\documentclass[abstract=true]{scrartcl}

\usepackage{graphicx}
\usepackage{amssymb}
\usepackage{amsmath}
\usepackage{natbib}
\usepackage{subcaption}

\usepackage[bookmarks,backref=false,linkcolor=black]{hyperref} 
\hypersetup{
  pdfauthor = {},
  pdftitle = {},
  pdfsubject = {},
  pdfkeywords = {},
  colorlinks=true,
  linkcolor= blue,
  citecolor= blue,
  pageanchor=true,
  urlcolor = blue,
  plainpages = false,
  linktocpage
}

\usepackage[capitalize,noabbrev]{cleveref}

\newcommand*{\doi}[1]{\href{https://dx.doi.org/\detokenize{#1}}{doi: \detokenize{#1}}}

\renewcommand{\cite}[1]{\citep{#1}}

\title{On Visualizing Data in\\Multi-display Environments}

\title{Multi-display Visual Analysis: Model, Interface, and Layout Computation}

\author{
C. Eichner \and
H. Schumann \and
C. Tominski}
\date{}
\publishers{
Institute for Visual \& Analytic Computing\\University of Rostock, Germany
}

\begin{document}
\maketitle

\begin{abstract}
Modern display environments offer great potential for involving multiple users in presentations, discussions, and data analysis sessions. By showing multiple views on multiple displays, information exchange can be improved, several perspectives on the data can be combined, and different analysis strategies can be pursued.

In this report, we describe concepts to support display composition, information distribution, and analysis coordination for visual data analysis in multi-display environments. In particular, a basic model for layout modeling is introduced, a graphical interface for interactive generation of the model is presented, and a layout mechanism is described that arranges multiple views on multiple displays automatically. Furthermore, approaches to meta-analysis will be discussed.
The developed approaches are demonstrated in a use case that focuses on parameter space analysis for the segmentation of time series data.

\end{abstract} 

\section{Introduction}
\label{sect:multi-display}

Typically visual analysis solutions are designed for a single user working with one or two displays. However, such environments are limited in two regards. First, only a single individual is involved in the data analysis. Critical reflections of results or creative discussions of alternative analysis strategies are hardly possible. Second, the available display space is limited. This can make the analysis of larger volumes of data difficult.

\index{display ecologies}
Addressing these two limitations, the goal of active research is to bring visual data analysis to advanced display environments that enable more data to be observed by more users. A natural step to achieve this goal is to combine several displays to form so-called \emph{display ecologies}~\cite{Chung15DisplayEcology}. Display ecologies can facilitate visual data analysis in different ways. The increased overall display space makes it possible to visualize not only more data, but also to look simultaneously at different aspects of the data. The increased physical size of the display space allows multiple users to study the data, which promotes collaborative analytic work.

While offering many advantages for visual data analysis, display ecologies also challenge users with additional tasks:

\begin{description}
	\item[Display composition] The heterogeneous displays of the environment must be integrated to form a coherent display space. This includes devices that are permanently connected to the display ecology, such as projectors or touch-sensitive surfaces, but also devices that enter or exit dynamically as users bring their own laptops, tablets, or smartphones.

	\item[Information distribution] The visual representations of the data and auxiliary information must be distributed in the environment. It must be decided, which display should show which information, and per display a suitable layout of the information must be created. The distribution must be flexible to allow new or updated information to be added to the environment dynamically.
	
	\item[Analysis coordination] The findings made on different displays by different users must be integrated to form a consistent overall picture of the analyzed data. This involves coordinating analysis sessions, discussing hypotheses, and managing intermediate results. In retrospect, it must be possible to understand how certain analysis steps contributed to the generation of new insight.
\end{description}

It is obvious that the above tasks would significantly increase the burden on the human user. Therefore, data analysis in display ecologies must be supported by appropriate methods that relieve users of laborious and time-consuming manual work and allow them to concentrate on their analytic objectives. This report briefly outlines how such support can be provided.

\subsubsection*{Scenario}

\index{smart environments}
For the purpose of illustrating display composition and information distribution, we consider a scenario where visualizations are presented and discussed in a smart meeting room. A \emph{smart meeting room} is an instance of smart environments, which Cook and Das define as follows~\cite{Cook04SmartEnvironments}:

\begin{quote}
	``A smart environment is a small world where all kinds of smart devices are continuously working to make inhabitants' lives more comfortable.''
\end{quote}

\begin{figure}
	\centering
	\includegraphics[width=\textwidth]{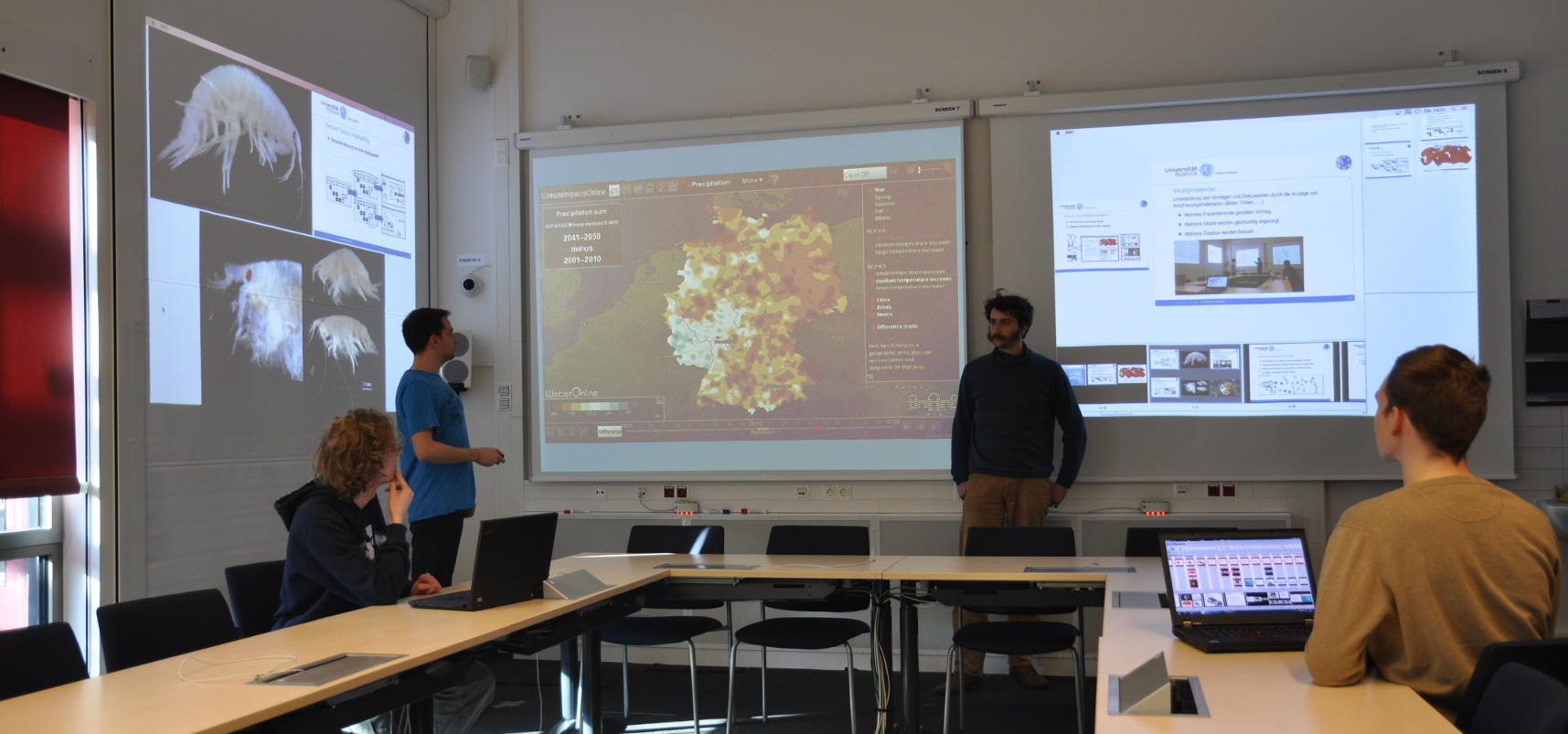}
	\caption{Smart meeting room at the University of Rostock. From~\cite{Eichner15DisplayEcologies}.}
	\label{fig:ch-advanced:smart-room}
\end{figure}

\cref{fig:ch-advanced:smart-room} shows an example of a smart meeting room. As can be seen, the room contains several devices for displaying information. What cannot be seen is that the room also integrates numerous sensors for tracking the environment and its inhabitants, and a pool of software tools for assisting the users. In particular, the room provides the technical basis for smoothly integrating different input, computing, and output devices into the device ensemble and for transferring information between multiple devices and displays. The room also evaluates its sensor data to recognize certain situations, reason about the users' intentions, and adapt the environment accordingly.

But how can a smart meeting room offer all these services? What is needed at the back-end and at the front-end to enable advanced visual analysis in smart meeting rooms?

In the first place, \emph{models} are needed. The smart room already has an internal model that describes the technical environment and the information about the users. On top of that, we need an appropriate model that formally describes advanced analysis sessions. The challenge is to cope with the highly dynamic character of interactive visual data analysis: Hypotheses can be subject to critical discussions leading to intermediate findings being revised and alternative courses of action being proposed on the fly.

In addition to models, suitable \emph{user interfaces} are needed to configure the involved models and adapt them dynamically, which effectively controls the state and the behavior of the room. This includes contributing content to be analyzed, declaring layout preferences, controlling the session progress, and returning to previously derived analysis results. The user interfaces must be prepared to deal with multiple users working together.

Finally, \emph{algorithms} are needed to drive the environment and provide assistance to the users. The smart meeting room already comes with algorithms to transfer data among devices and integrate displays to a coherent space. This way the technical basis for display composition and information distribution is provided. What has to be added are algorithms that distribute visual representations to multiple displays according to user-specified preferences. Moreover, we need algorithms that support the analysis coordination. For all tasks, it is necessary to employ efficient algorithms that can quickly adapt their results according to the dynamically changing situation in the room.

\bigskip

With the general context of our scenario being clear now, we can next go into the details about visual data analysis in smart rooms. In particular, there are three phases:

\begin{enumerate}
	\item Preparation: Several users plan and prepare an analysis session using existing visualization applications, images, reports, and other documents.
	\item Visual Analysis: An automatic layout algorithm implements the distribution and arrangement of visualizations across the displays. Then, the analysis starts. A moderator guides through the session, while all users can discuss prepared content and contribute additional content as necessary.
	\item Meta Analysis: The course of the analysis is automatically recorded. This allows decision makers, contributing authors, or persons from the general audience to review the analysis session for insight provenance.
\end{enumerate}

\section{Preparing Multi-display Analyses}

First, we consider the preparation phase. Basically, three questions have to be addressed:

\begin{itemize}
	\item What? Content.
	\item When? Sequence.
	\item Where? Display and layout.
\end{itemize}

The \emph{what} is about the content to be presented and discussed. Contents of different kinds can be contributed by any (authorized) user. Existing visual representations, pages from a report, or slides from a talk are examples of static content. Content can also be active. For example, a visualization tool can be linked with the content so that visual representations can be generated on demand.

The \emph{when} is about the sequence in which the content is to be presented. It defines the logical structure of the points to be communicated to the audience. The sequence of content is not set in stone. Quite the contrary, new or altered analysis objectives may change the time when content is presented.

Finally, the \emph{where} captures how the content is to be distributed in the environment and on the individual displays. This aspect would require particularly elaborate preparations if the users had to take on the task of defining the spatial arrangement manually. Yet, thanks to being in a smart room, a suitable spatial arrangement can be computed automatically. However, in light of a dynamically changing environment where the audience can bring their own devices and content, it is almost impossible to find a comprehensive solution in advance. Moreover, a purely automatic solution would prohibit the moderator from controlling the layout during the analysis session.

Therefore, it makes sense to handle the \emph{where} aspect semi-automatically in a mixed effort where human and computer complement each other. The users provide constraints to specify their preferences about where content should be displayed. The system evaluates these constraints and computes a spatial arrangement that suits both the situation of the environment and the users' needs.

The advantage of this approach is the following. The contributing users can focus on the logical communication of content, the \emph{what} and the \emph{when}. Their job of treating the \emph{where} aspect is eased considerably, because only qualitative constraints need to be declared, while the distribution of content and the precise specification of quantitative display positions is done by the system.

Next, the aforementioned aspects will be cast into an abstract model. This model provides the basis to plan, run, and steer visual analysis sessions in a multi-display environment.

\subsection{Abstract Model}

The content to be analyzed is collected in a \emph{content pool} $V$. For the sake of simplicity, we abstract from the concrete type of content and say that the content pool consists of views $v \in V$. A \emph{view} can be any \emph{static} visual representation or the output of an \emph{actively} executed visualization tool.

Given the content pool, the next question is in which order views should be presented. This can be modeled as a sequence of \emph{temporal layers}
\[
L = (L_1, L_2, \dots, L_n)
\]
where $L_i \subseteq V : 1 \leq i \leq n$ is the subset of the content pool to be shown at time step $t_i$. In other words, temporal layers structure a session in terms of what is shown when. The layers do not consider any aspects of spatial arrangement.

The spatial arrangement of content is to be derived by the system according to constraints defined by the users. Two types of constraints can be employed: spatial constraints and temporal constraints.

\emph{Spatial constraints} $C^S$ tell the system that certain views should be displayed close to each other. They are formally modeled as relations between views
\[
C^S = \bigcup_{i=1}^{n} C_i^S : C_i^S \subseteq (L_i \times L_i)
\]
Note that spatial constraints may only exist between views of the same layer $L_i$. Declaring spatial constraints is only a first option to control the automatic computation of suitable spatial arrangements.

In order to keep the arrangement reasonable stable over time, it makes sense to further specify the intended behavior when switching from one time step to the next. For this purpose, \emph{temporal constraints} can be used to tell the system which views of one time step are best to be replaced with which views of the following time step. Using a similar notation as before, temporal constraints $C^T$ are modeled as follows
\[
C^T = \bigcup_{i=1}^{n-1} C_i^T : C_i^T \subseteq (L_i \times L_{i+1})
\]
Note that temporal constraints may only exist between views of subsequent layers $L_i$ and $L_{i+1}$.

\begin{figure}
	\centering
	\includegraphics[width=\textwidth]{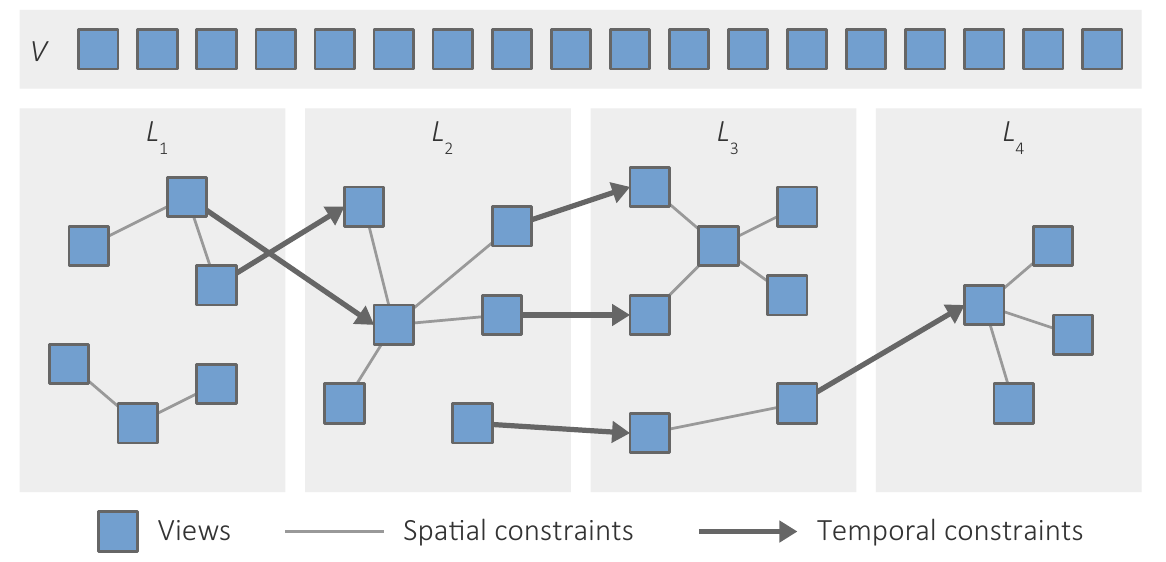}
	\caption{Abstract model of a multi-display visual analyis session.}
	\label{fig:ch-advanced:mde-model}
\end{figure}

\cref{fig:ch-advanced:mde-model} schematically summarizes the defined model with the content pool $V$, the temporal layers $L_i$, the spatial constraints $C^S$ defined within layers, and the temporal constraints $C^T$ defined between subsequent layers. The figure makes clear that the model corresponds to a layered graph. This graph enables the system to decide when and where visual content should be presented. On the other hand, the graph needs to be set up before the analysis and be adapted dynamically during the analysis according to the situation at hand. This requires a suitable graphical interface as described next.

\subsection{Graphical Interface}\label{sect:interface}

The graphical interface has to serve two purposes. First, it must afford creating and editing the model, which includes contributing views to the content pool, assigning views to layers, and defining spatial and temporal constraints between the views. Second, it should provide informative visual feedback about the model.

\begin{figure}
	\centering
	\includegraphics[width=\textwidth]{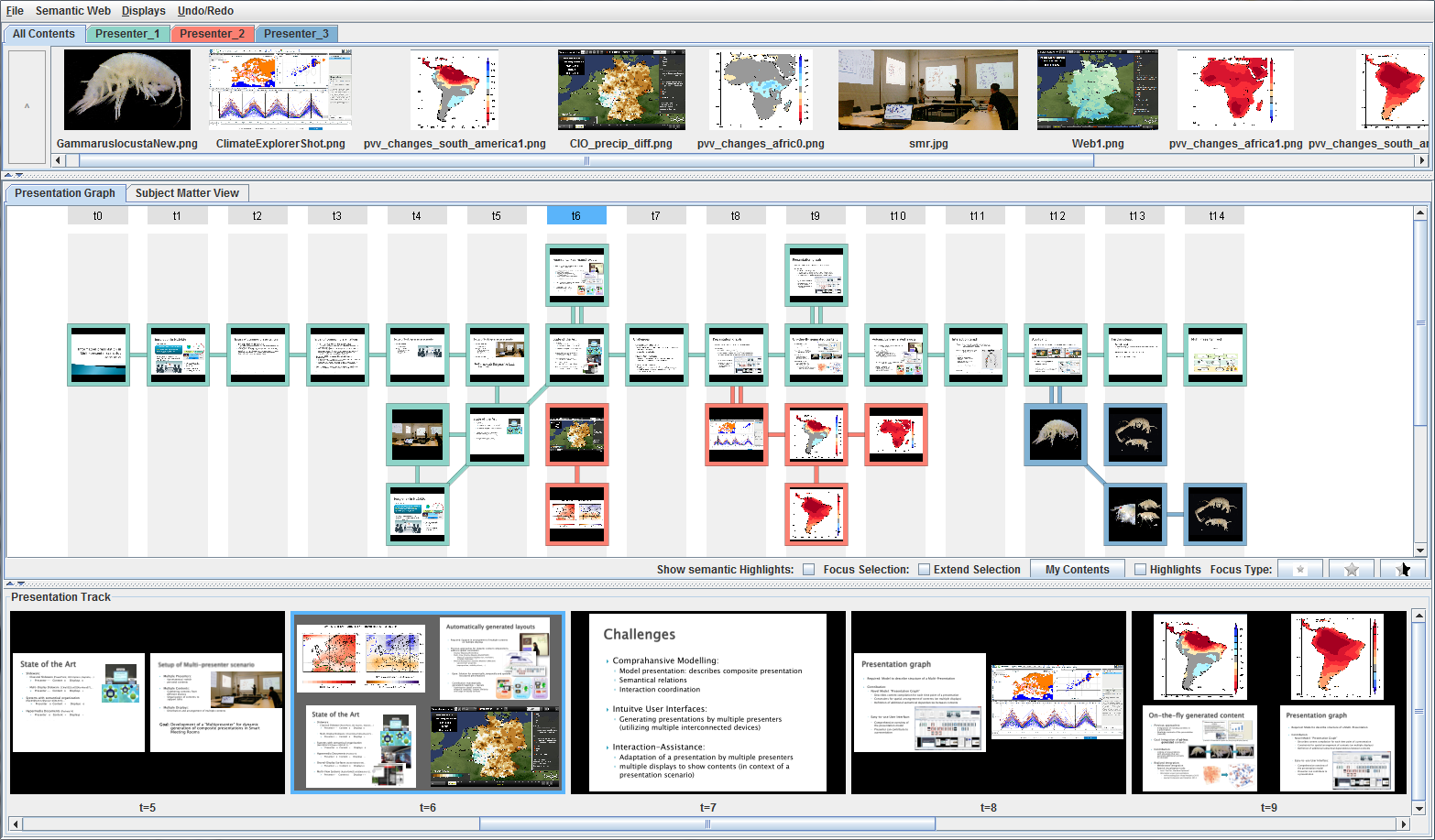}
	\caption{Graphical interface with content pool (top), logical session structure (middle), and preview (bottom) for preparing and controlling multi-display visual analysis sessions.}
	\label{fig:ch-present:graphical-interface}
\end{figure}

\cref{fig:ch-present:graphical-interface} shows a graphical interface that supports these tasks. The interface is available to all users in the smart meeting room, more precisely, each user has an own individual interface to access to the underlying model. When a user enters the smart meeting room and connects a personal device to the environment, the graphical interface will be shown on that device to enable the user to participate.

In the first place, users can contribute to the content pool $V$ shown in the top panel of \cref{fig:ch-present:graphical-interface}. Static content such as images, slides, or documents can simply be dragged and dropped to the content pool. Similarly, so-called \emph{active views} can be dropped into the content pool. An active view not only includes a visual representation, but also a link to the visualization software that was used to generate it. At any time, the linked software, provided that it is compatible, can be launched with a simple click to modify the existing visual representation or create a new one on the fly.

The middle panel of \cref{fig:ch-present:graphical-interface} serves to prepare the logical structure of the analysis session. The temporal layers $L_i$ are shown as columns. Views can be assigned to the layers by dragging them from the content pool onto the desired columns. Temporal and spatial constraints can be specified in a similar fashion by creating edges between views. For example, if a user wants a data table to be displayed next to a map visualization, the only thing he or she has to do is to draw an edge between the two thumbnails of the data table and the map. Colored frames and edges will indicate which user has made the corresponding edits.

The bottom panel in \cref{fig:ch-present:graphical-interface} shows a preview of the analysis session as defined by the layered graph. It tells the users which views are currently being shown in the smart meeting room, which views were displayed before, and which views are still to come. The user who moderates the analysis can advance the session to the next layer, return to the previous one, or override the step-wise progression and directly go to any layer if need be. Finally, the graphical interface allows views to be prioritized by assigning them different degrees of interest.

\bigskip

In summary, the described graphical interface enables multiple users to prepare a multi-display visual analysis session based on an underlying abstract model. A moderator chairs this process to ensure a consistent presentation that matches the objectives of the analysis session.

\section{Visual Analysis on Multiple Displays}

For the visual analysis, the modeled analysis session is executed in the smart meeting room. During the course of the session, the involved views are automatically distributed and arranged on the room's displays, and users interact with the system to adjust views and make progress in the analysis. Both aspects will be described next.

\subsection{Automatic View Layout}\label{sect:layout}

\index{multiple views}
While the smart room provides the technical basis for presenting views on multiple displays, a dedicated software component is in charge of calculating where exactly views are to be displayed. 
This view layout problem relates to the challenge of information distribution in display ecologies.
Here numerous influencing factors, including the physical properties of displays, the current situation in the room, the number, size, and importance of views, and the structure and constraints as specified by the underlying model must be considered.

\begin{figure}
	\centering
	\includegraphics[width=\textwidth]{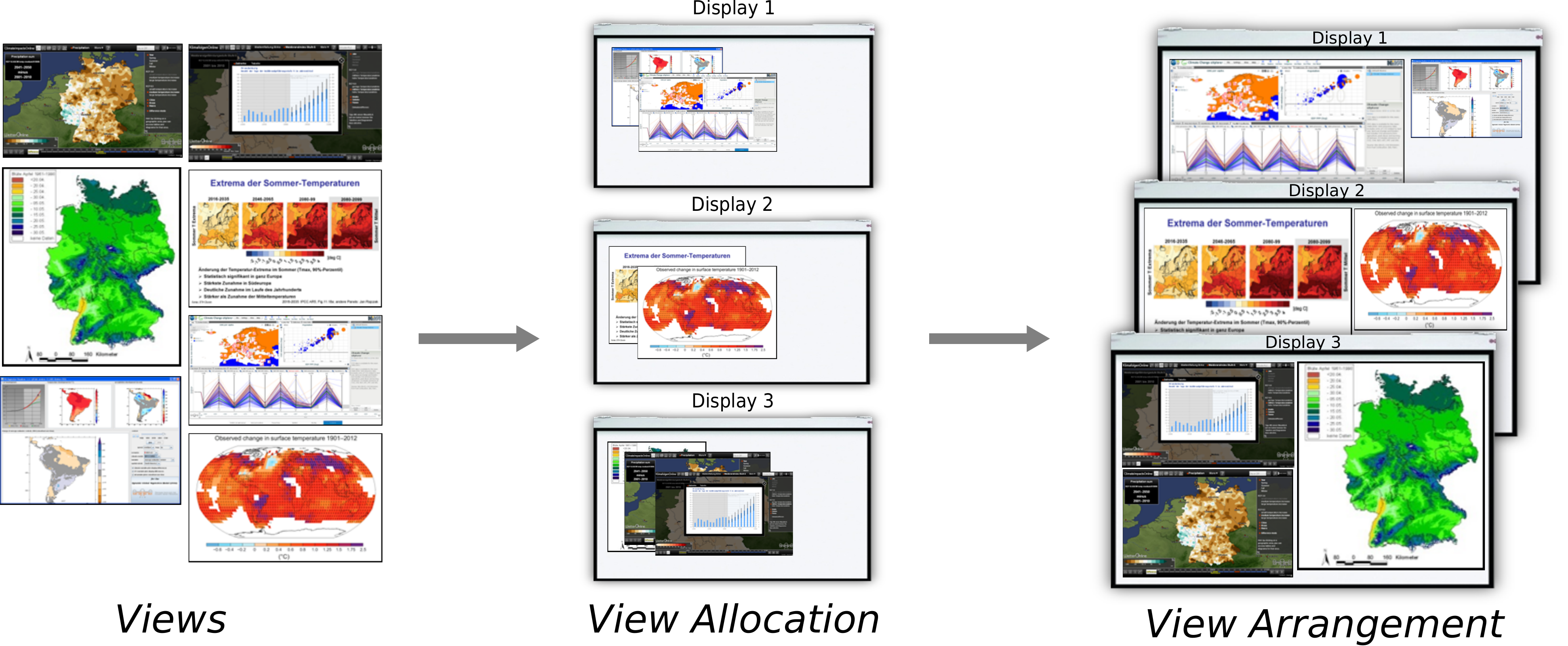}
	\caption{Two tasks for automatic view layout generation.}
	\label{fig:ch-present:automatic-layout}
\end{figure}

The layout problem can be divided into two aspects (cf. Figure~\ref{fig:ch-present:automatic-layout}).
Firstly, views must be assigned to the available displays and secondly, a suitable arrangement of views per display must be computed. 
Both aspects are not completely independent of each other: A poor allocation of views to displays, for example, can make it very difficult to find good layouts for individual displays.
On the other hand, good layouts don't do much good if they are only generated for displays outside the user's field of view.
Hence, view allocation and layout generation have to be considered together.

Under such complicated circumstances, it makes sense to define the automatic layout as an optimization problem~\cite{Eichner15Presentation}. Given a current time step $t_i$, the goal of the optimization is to find for each view $v \in L_i$ a position $p_v$ and a size $s_v$ such that the overall quality $Q$ with 
\[ Q = \alpha \cdot Q_S + \beta \cdot Q_T + \gamma \cdot Q_V \]
is maximal. The individual terms of $Q$ capture the spatial quality $Q_S$, the temporal quality $Q_T$, and the visibility quality $Q_V$ of a view.
The weights $\alpha$, $\beta$ and $\gamma$ can be adjusted to control the optimization, for example, to prioritize good visibility at the price of accepting compromises in terms of spatial or temporal quality.

The \emph{spatial quality} $Q_{S}$ is high if the views being linked via spatial constraints are indeed near to each other. This is modeled as:

\[ Q_S = \sum_{(u,v) \in C^S_i} 1 - \frac{\left|p_u-p_v\right|}{ext(D)} \]

As defined earlier, $C^{S}_{i} \subset C^{S}$ is the subset of spatial constraints associated with the time step $t_{i}$. $C^{S}_{i}$ consists of pairs of views $(u,v) : v, u \in L_{i}$ that are supposed to be displayed in close spatial proximity. The positions of $u$ and $v$ are denoted as $p_v$ and $p_u$, respectively. The extent $ext(D)$ of the display $D$ where $v$ and $u$ are shown determines the maximal possible distance between the views.

The \emph{temporal quality} $Q_{T}$ is high if temporally stable layouts are produced. That is, views being linked via temporal constraints $C^T_i$ are ideally presented at the same position when progressing from one time step to the next:

\[ Q_T = \sum_{(u,v) \in C^T_i} 1 - \frac{\left|p_u-p_v\right|}{ext(D)} \]

The \emph{visibility quality} $Q_V$ rates how well users can see the different views in the display environment. As a rule of thumb, views with a higher degree of interest, denoted as $doi(v)$, should exhibit a larger size, denoted as $s_v$. This leads to:

\[
Q_V = \sum_{v \in L_i} vis(v) \cdot \left(
        - doi(v) \cdot \left(\frac{s_v}{ext(D)}\right)^2
+ 2 \cdot doi(v) \cdot \left(\frac{s_v}{ext(D)}\right) \right)
\]

The governing factor $vis(v)$ captures the directional visibility of a views~\cite{Radloff11SmartViews}. The smart meeting room approximates this factor based on the display configuration (size, position, and orientation of displays) and the participating users (position and viewing direction of users).

In light of the above formulations it is clear that exhaustively testing all view positions and sizes during the optimization is impractical, if not impossible. 
Moreover, it can be necessary to resolve the optimization problem frequently, for example, when the degree of interest of views is adjusted or a mobile display is relocated, but also when the visibility of views changes as the moderator walks in front of the displays in the smart meeting room.

Approaches that try solve the optimization problem approximatively can cause problems if influencing factors such as the importance of views change.
For example, an approximatively found local quality maximum could be insensitive to minor adjustments.
This means small adjustments to the quality function initially have no effect at all and are then adopted suddenly if a better (local) quality maximum is found. This behavior would make layout adjustments unpredictable, often unstable, and difficult for users to control.
We therefore utilize an approach that seeks the absolute optimum of quality and thus reacts stably to all adjustments of the quality formula.

The view allocation, is done with a \emph{branch-and-cut} approach, which systematically generates and tests all possible view-display assignments.
For each view-display assignment the layouts are optimized on a per display basis to find the exact view position and size.
The individual layout problem is formulated as a quadratic optimization problem (with the target function $Q$) and restricted to consider only the views on the respective display. 
Additional linear contraints are added to prevent adjacent views from overlapping or protruding beyond the display boundary.
A Simplex algorithm can be applied to effectively calculate the optimal position and exact size of the views for one display.
The overall quality of a view-display assignment with all views and all displays results from the joint evaluation of all individually generated layouts.

In order to further accelerate the search for the best solution, the layout quality of a view-display assignment is (over-)estimated by an easy to calculate heuristic function.
In this way, promising view-display assignments can be considered early in the search and many low-quality assignments can be excluded from the exact layout calculation before the Simplex algorithm even starts.
In addition, partial solutions are reused if, for example, two different assignments require the same layout for a single display.
The described approach allows to calculate suitable and stable layouts with a dozen of views in about a second, which is totally fine for the addressed analysis scenario.

The automatic view layout is an enabling step for the visual data analysis. The users are not burdened with manually distributing and arranging views on multiple displays. It is merely necessary to specify a few constraints based on which the layout algorithm can operate.

\subsection{Visual Analysis and User Interaction}
\label{sec:inter}

Once the display environment shows the desired views, the data analysis can start. Led by a moderator, all people in the room can participate in the formulation of hypotheses, the discussion of findings, and the crystallization of insight. As this is a highly dynamic process, interaction with the views is important.

With changing topics of interest, the actual analysis situation might divert from the originally planned analysis session. Fine details spotted during the analysis could make it necessary to enlarge a view. Other views might need to be moved from one display to another for side-by-side comparison. Moreover, it may become necessary to change the views' content, for example, to show a different visual encoding, a different part of the data, or the same data at a different scale. The sketched operations suggest that two types of user interaction should be supported: adjustment of the view content and adjustment of the view layout.

\paragraph{Changing the Content of Views}

\begin{figure}
	\centering
	\includegraphics[width=\textwidth]{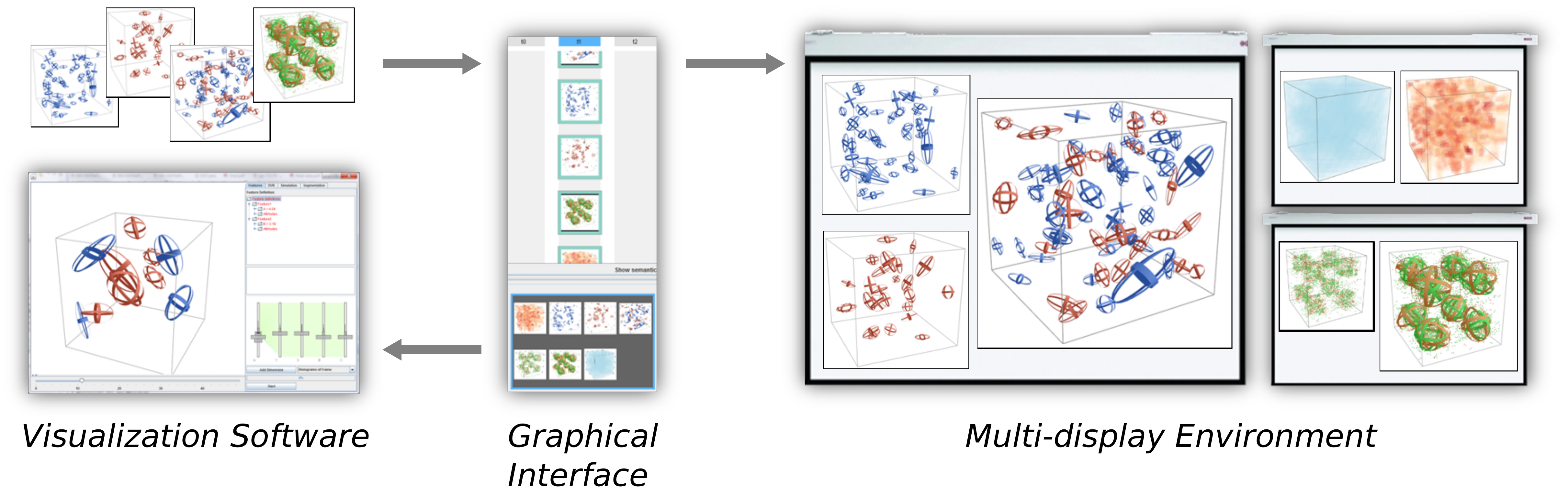}
	\caption{Changing the content of views by launching visualization software.}
	\label{fig:ch-advanced:adjust-view-content}
\end{figure}

We already mentioned that active views are linked with a compatible visualization software. This makes it possible to re-create the content of active views so that they better suit the task at hand. \cref{fig:ch-advanced:adjust-view-content} illustrates an example with the feature-based visualization tool described in \cite{Eichner14Feature}.

In order to change the content of a view, the linked visualization tool is launched from the graphical interface with a click on the view's thumbnail. Within the tool the existing content can be altered or a totally new visual representation can be generated. Once this is done, the new content is stored in the content pool and integrated into the model describing the logical structure of the analysis session. Then the displays of the environment are updated accordingly.

Typically, the new content of active views is generated locally on one of the personal devices connected to the smart environment. Hence, standard means of interaction are sufficient. On the other hand, adjusting the overall view layout requires dedicated means of interaction that operate in a unified interaction space. This will be explained next.

\paragraph{Adjusting the Layout of Views}

\begin{figure}
	\centering
	\begin{subfigure}[b]{0.325\textwidth}
		\includegraphics[width=\textwidth]{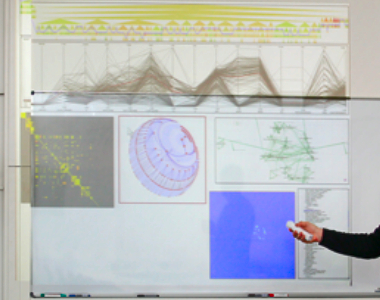}
		\caption{Point at central view.}
		\label{fig:ch-automatic:adjust-view-layout:a}
	\end{subfigure}
	\hfill
	\begin{subfigure}[b]{0.325\textwidth}
		\includegraphics[width=\textwidth]{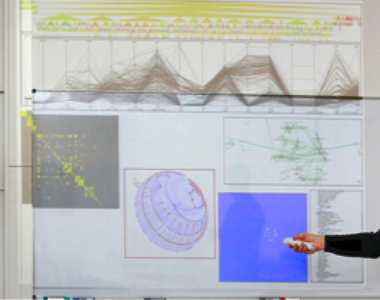}
		\caption{Move view downward.}
		\label{fig:ch-automatic:adjust-view-layout:b}
	\end{subfigure}
	\hfill
	\begin{subfigure}[b]{0.325\textwidth}
		\includegraphics[width=\textwidth]{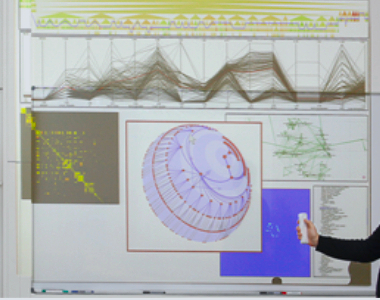}
		\caption{Enlarge view.}
		\label{fig:ch-automatic:adjust-view-layout:c}
	\end{subfigure}
	\caption{Adjusting position and size of a view using a Wii Remote controller. From \cite{Radloff12SmartInteraction}.}
	\label{fig:ch-advanced:adjust-view-layout}
\end{figure}

Here, we consider an example with visual representations of a graph as generated by the CGV system~\cite{Tominski09CGV}. The example in \cref{fig:ch-advanced:adjust-view-layout} now shows the graph visualizations projected onto the wall of the smart meeting room. The moderator points at the central Magic Eye View, moves it a bit downward, and then enlarges the view to make it stand out.

From a user's perspective, these operations are quite easy to perform with a Wii Remote controller. Yet, to make these seemingly simple adjustments of the view layout possible, dedicated mechanisms have to be implemented on the system's side, including an interaction grabber, an interaction mapper, and an interaction handler~\cite{Radloff15VisualComputer}.

There are many different ways how users can interact in a multi-display environment, for example, with classic mouse and keyboard interaction as well as modern touch-based or tracking-based interactions. The \emph{interaction grabber} collects all interaction events and converts them to a generic format to support fundamental pointing and triggering. The task of the \emph{interaction mapper} is to determine the display where an interaction is to take effect and to delegate the interaction request to the computing device that is responsible for handling it. Finally, the \emph{interaction handler} interprets the interaction, executes the necessary changes, and notifies the system of the update.

In our example, the interaction grabber gathers events from a Wii Remote controller held by the discussion moderator. The interaction mapping allows the moderator to point at any display connected with the smart room. Using different gestures and the controller buttons, the moderator can select views, relocate them, or adjusts their size, as already seen in \cref{fig:ch-advanced:adjust-view-layout}. Any user can perform these and other interactions via their personal interaction devices and the graphical interface. Yet, the moderator should guide and coordinate the interactions to avoid or resolve conflicting actions.

Next, we will explain that all interactions are logged on a per-user basis, not only for undo and redo, but also for a meta analysis of the actual visual data analysis.

\subsection{Coordination and Meta Analysis}\label{sect4:MetaAnalysis}

When multiple users engage in multi-display data analysis activities, coordinating the insight-generation process and reflecting about it can become a challenge. This section briefly explores how analysts and decision makers can be supported based on information recorded and annotated during the visual analysis.

To be able to coordinate and understand the visual analysis, it is necessary collect information about it:

\begin{itemize}
	\item What types of interactions were performed?
	\item Who initiated the interactions?
	\item Where have certain views been displayed?
	\item When were views visible?
	\item What findings could be derived from views?
\end{itemize}

Some of this information can be determined automatically by the smart meeting room. For example, while the automatic view layout is doing its work, the system automatically keeps track of when and where views were shown. When the layout or the content of views is changed interactively, the smart room automatically logs not only what actions were taken, but also who carried them out.

Yet, some information cannot be derived automatically. For example, when users spot something interesting, the corresponding views need to be annotated manually to document the finding. Both, the automatically recorded and the manually annotated information for each view is stored in an analysis log.

\begin{figure}
	\centering
	\includegraphics[width=.9\textwidth]{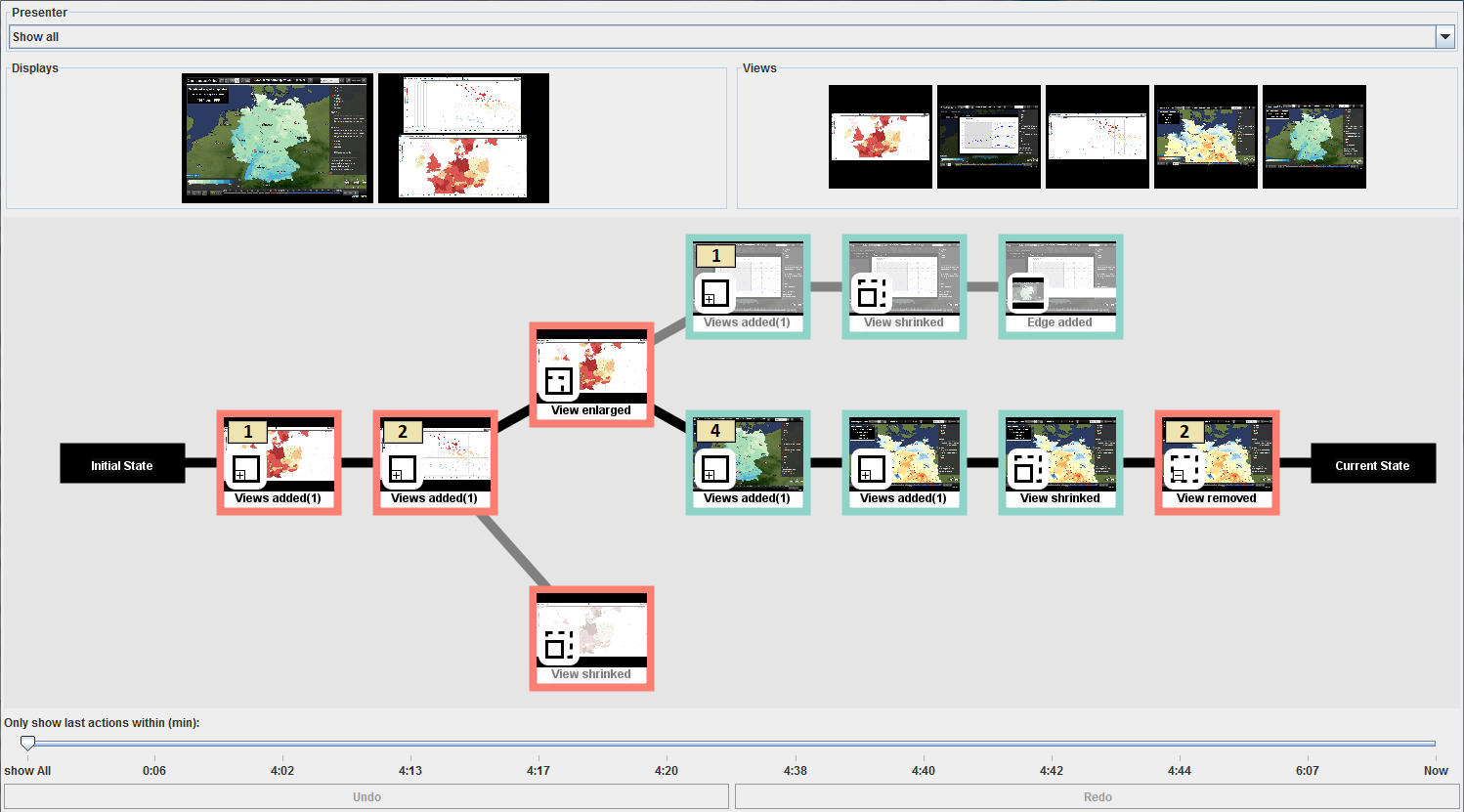}
	\caption{Graphical interface for analysis coordination and meta analysis. 
	}
	\label{fig:ch-advanced:interaction-graph}
\end{figure}

The analysis log forms a graph that serves as the basis for coordination and meta analysis. The graph's nodes represent views, more precisely, the state changes logged per view. In a sense, a node captures a piece of analytical progress made during the data analysis. Links between nodes form paths of analytical progress as defined by the sequence of actions stored in the analysis log. To make the analysis log accessible to users, nodes and links are visualized in a dedicated graphical interface. A small example is provided in \cref{fig:ch-advanced:interaction-graph}. This interface can be utilized during the visual analysis for coordination and afterwards for a meta analysis.

During the visual analysis, undo and redo operations can be performed, that is, the analysis can be reset to a previous state. This is helpful when the data analysis stalls in a dead end or if the participating users cannot come to an agreement about findings and intermediate analysis results. Undo and redo allows the moderator to keep the analysis going, for example, by collecting further evidence for or against a hypothesis from previous views. Note that the underlying analysis log enables a selective undo and redo. This makes it possible to restrict undo and redo to state changes that were triggered by a certain user, affected a specific view, or concerned a particular display.

If, after returning to a previous state, an alternative course of actions is pursued, a new analysis branch is created, which is also apparent in \cref{fig:ch-advanced:interaction-graph}. Inspecting the different outcomes of such alternative analysis paths can be part of a post-hoc meta analysis.

The goal of the meta analysis is to understand how certain analysis steps contributed to the generation of new insights. Again, the visualization of the branching graph of state changes in \cref{fig:ch-advanced:interaction-graph} plays a central role. In our case, three alternative analysis routes were tried out. Small icons overlaid on the thumbnails of the modified views indicate which interactions were performed, and the thumbnails' colored borders tell indicate who performed them. Our small analysis session involved only two users (green and red).

Additional information can be queried from the analysis log on demand. Clicking on a view will show a text box that informs the user about when and where the view was displayed, and which findings could be derived from it. Additional controls in \cref{fig:ch-advanced:interaction-graph} facilitate filtering larger analysis logs with respect to user, display, type of interaction, time, and findings. This way, the meta analysis can be narrowed down on particular questions of interest, such as who contributed most to the creation and adjustment of views, which interactions led to promising analysis paths, or which analysis results required a longer time of discussion before they were agreed upon.

\section{Application to Visual Parameter Space Analysis}

The presented multi-display visualization approach has already been applied in the context of expert discussions of climate data~\cite{Eichner15Presentation}.
In the following, we describe a use case that revolves around parameter space analysis for the segmentation of time series.

\subsection{Time Series Segmentation and Parameter Space Analysis}

Segmented time series are relevant in many data analysis scenarios, because they can summarize important processes of the original data.
For example, algorithms for activity detection rely on segmented time series from multiple sensors in order to draw conclusions about the activities of people.
To generate segmented time series, the raw data are processed by a segmentation pipeline~\cite{Bernard18Combining}. After an appropriate data pre-processing, a segmentation algorithm divides the time axis into several \emph{segments}. Finally, a labeling step annotates the generated segments with \emph{labels}.

Segmentation algorithms must be configured appropriately to deliver the desired segmentation results. For this purpose, various parameters can be tuned to influence the processing of the data and the segmentation outcome. However, it is not always clear how strongly and in what way parameters influence the segmentation. Therefore, the goal of a parameter space analysis is to determine the dependencies between parameters and segmentations.

A comprehensive analysis requires investigating influences of different parameters on different labels and to compare them with each other.
For that, overview representations of parameter-segmentation dependencies have to be considered in concert with visual representations of the generated segmented time series.
The analysis must also account for the fact that parameters can influence different properties of the segments and that dependencies may exists only in certain parameter ranges. Therefore, the parameter space analysis involves several kinds of visualizations to be interpreted in concert.

A multi-display environment is an excellent match for supporting this type of analysis, as the available display area on multiple displays can be used to show precise pixel-based representations in full resolution and to display multiple views in combination.
Next, we illustrate how an interactive analysis of parameter-segmentation dependencies and segmented time-series data can be carried out in a smart meeting room using our approach. The segmented time series represent human activities recognized from sensor data and the overall segmentation procedure uses five parameters in total.

\subsection{Multi-display Visual Parameter Space Analysis}

The analysis addresses two objectives: finding out what properties of the segmentation are influenced and determining parameters that are exerting influence~\cite{Eichner2019Correlations}. Here, we describe a parameter-first analysis strategy, for which the starting point are the parameters. A parameter-first analysis aims to find parameters that most likely have a strong influence on the segmentation.

The analysis is based on various dedicated visual representations. These visual representations are generated by a visualization tool and are distributed as \emph{active views} via the software infrastructure in the smart meeting room, where they automatically appear in the graphical user interface as discussed in \cref{sect:interface}.

In a first step, overview visualizations are used that show different parameterizations and corresponding segmentation results as colored pixel rows stacked on top of each other \cite{Roehlig15SupportingAR}.
By sorting the rows in different ways, dependencies between parameter values and segment properties can be investigated.
In order to estimate the influence of each parameter individually, five overviews are generated, each sorted with respect to a different parameter.
In addition, a dedicated view shows the variation of parameter influences. This view is based on calculated correlations, more precisely on deviations of correlations computed for parameters and features of the segmented time series~\cite{Eichner2019Correlations}.

\begin{figure}
	\includegraphics[width=\textwidth]{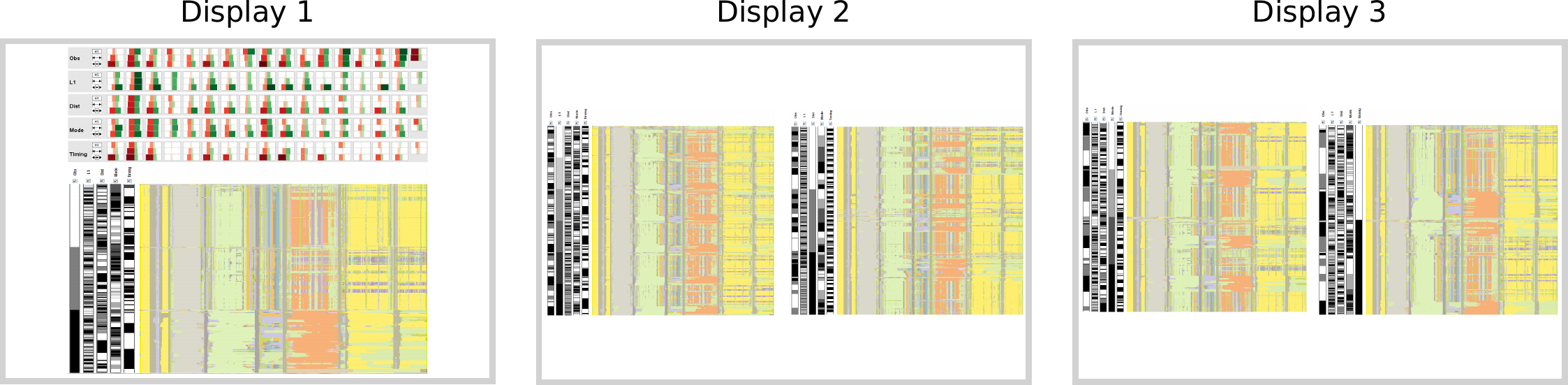}
	\caption{First step in the analysis session supported by six views on three displays. Five pixel-based overview visualizations show the segmentation data sorted with respect to different parameters. The sixth view shows deviations in the correlation of different combinations of parameters and segment properties.}
	\label{fig:Displays1}
\end{figure}

The six views are automatically distributed to three displays by the automatic view layout, as depicted in Figure~\ref{fig:Displays1}.
Showing the views on three displays makes it possible to get an overview, and also to inspect details. All users being present in the smart meeting room can participate in the analysis.
By comparing the differently sorted overviews and the information presented in the correlation view, the users can develop hypotheses on which parameters might be influential and which might be not. Subsequently, the analysis will focus on parameters with the strongest influence.

\begin{figure}
	\includegraphics[width=\textwidth]{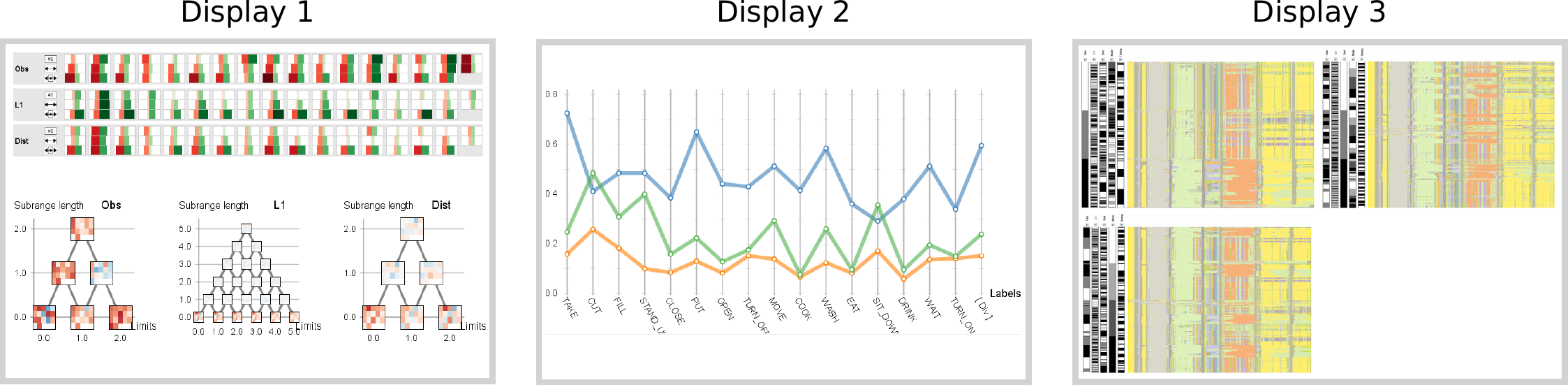}
	\caption{The second analysis step focuses on three influential parameters. The pixel-based overviews have been moved to Display 3. New triangular views are added to visualize the parameter influence in parameter subranges on Display 1 and to support comparison of parameter influence with a parallel coordinates plot on Display 2.}
	\label{fig:Displays2}
\end{figure}

In a second step, the analysts can investigate how strong the influence of the different parameters is and what properties of the segmented time series are influenced. This involves global influence for the entire range of parameter values and also local influence in parameter sub-ranges.
To carry out the second analysis step, the view composition in the smart meeting room is adjusted as described in Section~\ref{sec:inter}.
First, the overviews of high-influence parameters are placed on the same display, and the overviews sorted according to parameters with little influence are removed so that more display space is available (see Figure~\ref{fig:Displays2}). 
Additionally, for each important parameter, a triangular subrange visualization of correlation values is called up as a new view to help analysts assess the influence in different parameter sub-ranges \cite{Eichner2019Correlations}.
Finally, a parallel coordinate plot is added to support the comparison of the average correlation strength of the three remaining parameters for segments with different labels.
The comparison is easier when the visualizations to be compared are placed spatially close to each other. Therefore, the user interface is employed to define spatial constraints for the three overviews and the three triangular subrange visualizations. This leads the automatic layout computation to automatically group the views on Displays 1 and Display 3 as depicted in Figure~\ref{fig:Displays2}.
Moreover, the parallel coordinates plot is assigned a high degree of interest so that this view appears in full size on Display 2.
By investigating the new views for the second analysis step, users can examine in more detail how individual parameters exert their influence and more directly compare the parameter influence.

\begin{figure}
	\includegraphics[width=\textwidth]{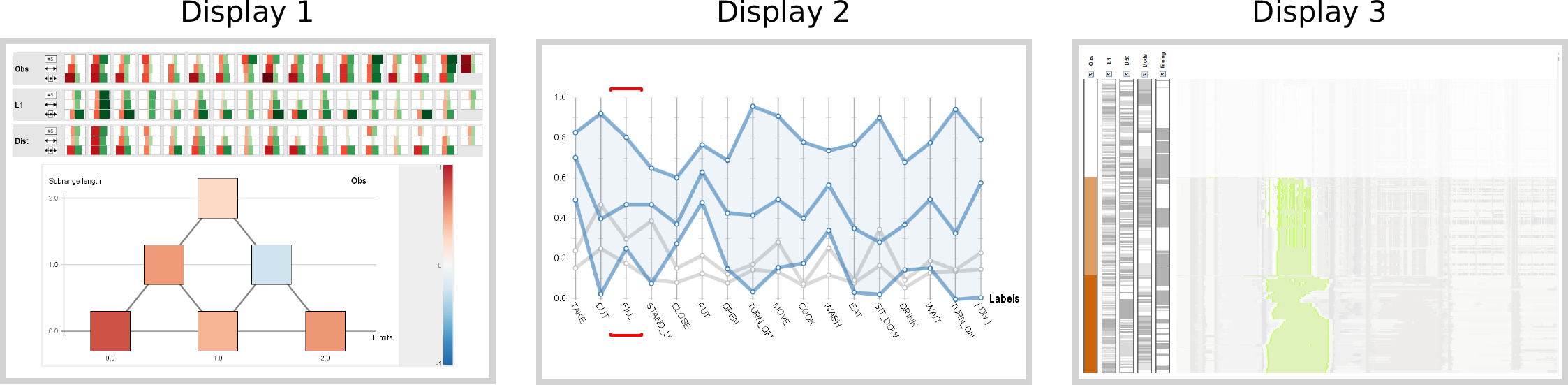}
	\caption{Third analysis step, focusing on a dependency between a selected parameter and a particular segmentation property. Irrelevant views are removed from the displays and the content of the remaining views is adapted to emphasize the investigated parameter-segment dependency.}
	\label{fig:Displays3}
\end{figure}

The third step of the analysis, aims to investigate the dependencies between a selected parameter and a particular segment property in more detail. This third step ultimately allows analysts to find suitable parameter values for the segmentation algorithm.
To this end, as before, views that do not contribute to the new analysis goal are removed. The content of the remaining views is automatically adjusted to make details of the investigated data more visible as in Figure~\ref{fig:Displays3}.
For example, the triangular subrange visualization on Display 1 is adjusted to show only the correlation values between a selected parameter and a certain segment property. The parallel coordinate plot on Display 2 emphasizes these correlation values separately for different labels and the pixel-based overview on Display 3 is changed to highlight the affected segments and the influencing parameter.

\begin{figure}
	\includegraphics[width=\textwidth]{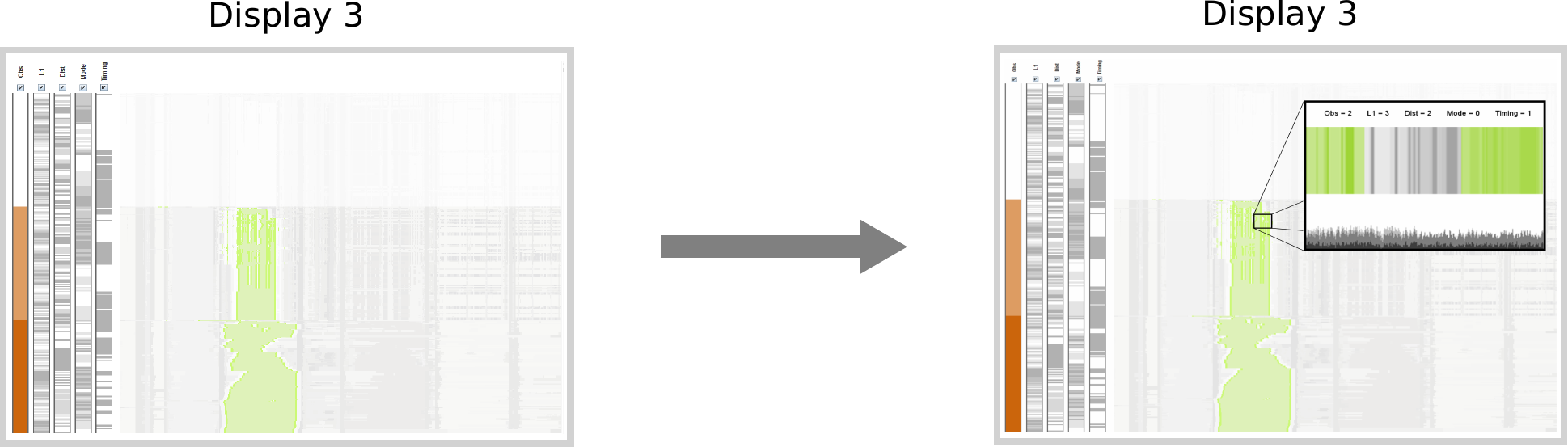}
	\caption{Application of an interactive lens on a pixel-based overview to make further details visible. The lens magnifies a temporal section and visualizes uncertainties of the label assignment and uncertainties introduced by sensor noise.}
	\label{fig:Displays4}
\end{figure}

Now, interactive lenses can be integrated as additional views to make further details of the data visible. Figure~\ref{fig:Displays4} shows the application of such a lens on an overview of the segmented time series. In this example, the lens allows analysts to examine uncertainties in the data, and thus, to take uncertainties into account as another analysis aspect when evaluating parameter influence.

By examining multiple views in concert, potential dependencies between individual parameters and certain segment properties can be evaluated. 
However, in order to investigate another dependency, the display would first have to be reset to the state for the second analysis step.
Instead of reverting each adjustment manually, the analysts use the meta-analysis interface presented in section~\ref{sect4:MetaAnalysis}.
It shows all adjustments made to the display so far and allows to easily revert the adjustments of the third analysis step.
Now the third analysis step can be repeated for a dependency between a different parameter or another type of segment.
Each of these analysis directions appears as a separate analysis path in the interface for the meta-analysis and can be easily restored for later review.

\section{Summary}

In conclusion, we see that bringing interactive visual data analysis to multi-display environments is an exciting opportunity for collaborative sense-making. Yet, before this opportunity can be fully exploited, several challenges have to be addressed: the challenge of display composition, the challenge of information distribution, and the challenges of analysis coordination. 

Here, we illustrated how these challenges can be tackled with a mix of automatic methods and interactive graphical interfaces, which together form an advanced visualization environment. We introduced an abstract model that allows users to specify spatial and temporal constraints for distributing multiple views in a dynamically changing display environment.
We described a graphical interface that allows users to edit the model and adjust the information distribution as necessary.
An automatic view layout algorithm is utilized to assign views to displays and to arrange multiple views automatically according to the model's constraints.
On top of that, approaches for the visual analysis are introduced to enable users to interactively change generated layouts and contents according to their needs.
To support meta-analysis, a graphical interface has been developed, that enables analysts to review the process of analysis sessions and to restore individual steps with only little effort.
The interplay of the newly developed approaches has finally been demonstrated in a use case centered around a visual parameter space analysis for time series segmentation.

\bibliographystyle{plainnat}

\bibliography{references}

\begin{thebibliography}{12}
\providecommand{\natexlab}[1]{#1}
\providecommand{\url}[1]{\texttt{#1}}
\expandafter\ifx\csname urlstyle\endcsname\relax
  \providecommand{\doi}[1]{doi: #1}\else
  \providecommand{\doi}{doi: \begingroup \urlstyle{rm}\Url}\fi

\bibitem[Bernard et~al.(2018)Bernard, Bors, B\"ogl, Eichner, Gschwandtner,
  Miksch, Schumann, and Kohlhammer]{Bernard18Combining}
J\"urgen Bernard, Christian Bors, Markus B\"ogl, Christian Eichner, Theresia
  Gschwandtner, Silvia Miksch, Heidrun Schumann, and J\"orn Kohlhammer.
\newblock {Combining the Automated Segmentation and Visual Analysis of
  Multivariate Time Series}.
\newblock In \emph{Proceedings of the EuroVis Workshop on Visual Analytics
  (EuroVA)}, 2018.
\newblock \doi{10.2312/eurova.20181112}.

\bibitem[Chung et~al.(2015)Chung, North, Joshi, and
  Chen]{Chung15DisplayEcology}
Haeyong Chung, Chris North, Sarang Joshi, and Jian Chen.
\newblock {Four Considerations for Supporting Visual Analysis in Display
  Ecologies}.
\newblock In \emph{Proceedings of the IEEE Conference on Visual Analytics
  Science and Technology (VAST)}, pages 33--40. IEEE Computer Society, 2015.
\newblock \doi{10.1109/VAST.2015.7347628}.

\bibitem[Cook and Das(2004)]{Cook04SmartEnvironments}
Diane Cook and Sajal~K. Das.
\newblock \emph{{Smart Environments: Technology, Protocols and Applications}}.
\newblock Wiley-Interscience, 2004.
\newblock \doi{10.1002/047168659X}.

\bibitem[Eichner et~al.(2014)Eichner, Bittig, Schumann, and
  Tominski]{Eichner14Feature}
Christian Eichner, Arne Bittig, Heidrun Schumann, and Christian Tominski.
\newblock {Analyzing Simulations of Biochemical Systems with Feature-Based
  Visual Analytics.}
\newblock \emph{Computers \& Graphics}, 38\penalty0 (1):\penalty0 18--26, 2014.
\newblock \doi{10.1016/j.cag.2013.09.001}.

\bibitem[Eichner et~al.(2015{\natexlab{a}})Eichner, Nocke, Schulz, and
  Schumann]{Eichner15Presentation}
Christian Eichner, Thomas Nocke, Hans~Jörg Schulz, and Heidrun Schumann.
\newblock {Interactive Presentation of Geo-Spatial Climate Data in
  Multi-Display Environments}.
\newblock \emph{ISPRS International Journal of Geo-Information}, 4\penalty0
  (2):\penalty0 493--514, 2015{\natexlab{a}}.
\newblock \doi{10.3390/ijgi4020493}.

\bibitem[Eichner et~al.(2015{\natexlab{b}})Eichner, Nyolt, and
  Schumann]{Eichner15DisplayEcologies}
Christian Eichner, Martin Nyolt, and Heidrun Schumann.
\newblock {A Novel Infrastructure for Supporting Display Ecologies}.
\newblock In \emph{Advances in Visual Computing: Proceedings of the
  International Symposium on Visual Computing (ISVC)}, pages 722--732.
  Springer, 2015{\natexlab{b}}.
\newblock \doi{10.1007/978-3-319-27863-6\_68}.

\bibitem[Eichner et~al.(2019)Eichner, Schumann, and
  Tominski]{Eichner2019Correlations}
Christian Eichner, Heidrun Schumann, and Christian Tominski.
\newblock {Making Parameter Dependencies of Time-Series Segmentation Visually
  Understandable}.
\newblock \emph{Computer Graphics Forum}, 2019.
\newblock \doi{10.1111/cgf.13894}.
\newblock to appear.

\bibitem[Radloff et~al.(2011)Radloff, Luboschik, and
  Schumann]{Radloff11SmartViews}
Axel Radloff, Martin Luboschik, and Heidrun Schumann.
\newblock {Smart Views in Smart Environments}.
\newblock In \emph{Proceedings of the Smart Graphics}, pages 1--12. Springer,
  2011.
\newblock \doi{10.1007/978-3-642-22571-0_1}.

\bibitem[Radloff et~al.(2012)Radloff, Lehmann, Staadt, and
  Schumann]{Radloff12SmartInteraction}
Axel Radloff, Anke Lehmann, Oliver~G. Staadt, and Heidrun Schumann.
\newblock {Smart Interaction Management: An Interaction Approach for Smart
  Meeting Rooms}.
\newblock In \emph{Proceedings of the Eighth International Conference on
  Intelligent Environments (IE)}, pages 228--235. IEEE Computer Society, 2012.
\newblock \doi{10.1109/IE.2012.34}.

\bibitem[Radloff et~al.(2015)Radloff, Tominski, Nocke, and
  Schumann]{Radloff15VisualComputer}
Axel Radloff, Christian Tominski, Thomas Nocke, and Heidrun Schumann.
\newblock {Supporting Presentation and Discussion of Visualization Results in
  Smart Meeting Rooms}.
\newblock \emph{The Visual Computer}, 31\penalty0 (9):\penalty0 1271--1286,
  2015.
\newblock \doi{10.1007/s00371-014-1010-x}.

\bibitem[R\"ohlig et~al.(2015)R\"ohlig, Luboschik, Kr\"uger, Kirste, Schumann,
  B\"ogl, Alsallakh, and Miksch]{Roehlig15SupportingAR}
M.~R\"ohlig, M.~Luboschik, F.~Kr\"uger, T.~Kirste, H.~Schumann, M.~B\"ogl,
  B.~Alsallakh, and S.~Miksch.
\newblock {Supporting Activity Recognition by Visual Analytics}.
\newblock In \emph{Proceedings of the IEEE Conference on Visual Analytics
  Science and Technology (VAST)}, pages 41--48. IEEE Computer Society, 2015.
\newblock \doi{10.1109/VAST.2015.7347629}.

\bibitem[Tominski et~al.(2009)Tominski, Abello, and Schumann]{Tominski09CGV}
Christian Tominski, James Abello, and Heidrun Schumann.
\newblock {CGV -- An Interactive Graph Visualization System}.
\newblock \emph{Computers {\&} Graphics}, 33\penalty0 (6):\penalty0 660--678,
  2009.
\newblock \doi{10.1016/j.cag.2009.06.002}.

\end{thebibliography}
\end{document}